\documentclass{ws-mplb}

\newcommand{\bn}{\begin{enumerate}}
\newcommand{\en}{\end{enumerate}}
\newcommand{\ba}{\begin{eqnarray}}
\newcommand{\ea}{\end{eqnarray}}

\usepackage{graphicx,color}

\newcommand{\be}{\begin{equation}}
\newcommand{\ee}{\end{equation}}

\newcommand{\ra}{\rangle}
\newcommand{\et}{{\it et al. }}
\newcommand{\ete}{{\it et al.}}

\def\prl{{ Phys. Rev. Lett. }}

\def\apl{{ Appl. Phys. Lett. }}
\def\prb{{ Phys. Rev. B }}

\newcommand{\gd}[3]{Gd$_{#1}$Fe$_{#2}$Co$_{#3}$}

\newcommand{\tb}[2]{Tb$_{#1}$Co$_{#2}$}
\newcommand{\tbfe}[2]{Tb$_{#1}$Fe$_{#2}$}

\newenvironment{roof}[2]
{\sqrt{\frac{#1}{#2}}
}

\begin{document}

\markboth{Zhang, Latta, Babyak, Bai and George }{All-optical spin switching:
 A short review and
 a simple theory}

\catchline{}{}{}{}{}
%
%%%%%%%%%%%%%%%%%%%%%%%%%%%%%%%%%%%%%%%%%%%%%%%%%%%%%%%%%%%%%%%%%%%%

\title{
 ALL-OPTICAL SPIN SWITCHING: \\ A NEW FRONTIER IN FEMTOMAGNETISM\\
 A SHORT REVIEW AND
 A SIMPLE THEORY
%\footnote{For the title, try not to
%use more than 3 lines. Typeset the title in 10 pt
%Times roman, uppercase and boldface.}
}

\author{\footnotesize G. P. Zhang$^*$, T. Latta, and Z. Babyak
%\footnote{Typeset names in
%10~pt Times roman, uppercase. Use the footnote to indicate
%the present or permanent address of the author.}
}

\address{Department of Physics, Indiana State University,
   Terre Haute, IN 47809, USA
%\footnote{State completely without abbreviations, the
%affiliation and mailing address, including country. Typeset in 8~pt
%Times italic.}\\
$^*$gpzhang@indstate.edu}

\author{Y. H. Bai}

\address{Office of Information Technology, Indiana State
  University, Terre Haute, IN 47809, USA}

\author{Thomas F. George}

\address{Office of the Chancellor and Center for Nanoscience
  \\Departments of Chemistry \& Biochemistry and Physics \& Astronomy
  \\University of Missouri-St. Louis, St.  Louis, MO 63121, USA }

\maketitle

\begin{history}
\received{(11 May 2016)}
%\published{(Day Month Year)}
\end{history}

\begin{abstract}
{Using an ultrafast laser pulse to manipulate the spin degree of
  freedom has broad technological appeal. It allows one to control the
  spin dynamics on a femtosecond time scale.  The discipline, commonly
  called femtomagnetism, started with the pioneering experiment by
  Beaurepaire and coworkers in 1996, who showed subpicosecond
  demagnetization occurs in magnetic Ni thin films. This finding has
  motivated extensive research worldwide.  All-optical
  helicity-dependent spin switching (AOS) represents a new frontier in
  femtomagnetism, where a single ultrafast laser pulse or multiple
  pulses can permanently switch spin without any assistance from a
  magnetic field. This review summarizes some of the crucial aspects
  of this new discipline: key experimental findings, leading
  mechanisms, controversial issues, and possible future
  directions. The emphasis is on our latest investigation. We first
  develop the all-optical spin switching rule that determines how the
  switchability depends on the light helicity. This rule allows one to
  understand microscopically how the spin is reversed and why the
  circularly polarized light appears more powerful than the linearly
  polarized light.  Then we invoke our latest spin-orbit coupled
  harmonic oscillator model to simulate single spin reversal. We
  consider both cw excitation and pulsed laser excitation.  The
  results are in a good agreement with the experimental
  result.\footnote{A MatLab code is available upon request from the
    authors.}  We then extend the code to include the exchange
  interaction among different spin sites. We show where the ``inverse
  Faraday field'' comes from and how the laser affects the spin
  reversal nonlinearly. Our hope is that this review will motivate new
  experimental and theoretical investigations and discussions. }
\end{abstract}

\keywords{All-optical, spin switching, ultrafast}

\section{Introduction}

In the course of magnetic recording, the types of magnetic materials
and how one manipulates the spin moment are critical to the
information technology industry. In normal magnetic storage media, the
information is stored in magnetic domains of a few microns or
less, with spin pointing in one direction called bit ``1,'' and the
other direction called ``0.'' The external reading/writing head
generates a small magnetic field, which switches those bits. Such a
mechanism is used in normal hard drives. In a magneto-optical recording,
one uses a laser beam to warm up the medium above the compensation
point; when an external magnetic field is applied, the magnetic bit is
switched over. However, there are some exceptions where a magnetic
field is not used. In 1985, Shieh and Kryder\cite{shieh} showed that
in a GdTbCo film, the demagnetizing field\cite{aharoni} can effectively
be used as a bias field for thermomagnetic writing without a magnetic
field. This also works for GdTbCo, TbCo and TbFeCo thin films of
several thousand angstroms thick. In their picture, the
demagnetization field acts as a switching kernel.  To understand how
microscopically the switching process happens, the real time-dependent
simulation of magnetic domain motion has been carried
out.\cite{aharoni} This is a pre-femtomagnetism era, where the
switching speed was not a major concern.

\begin{figure}[th]
\centerline{\psfig{file=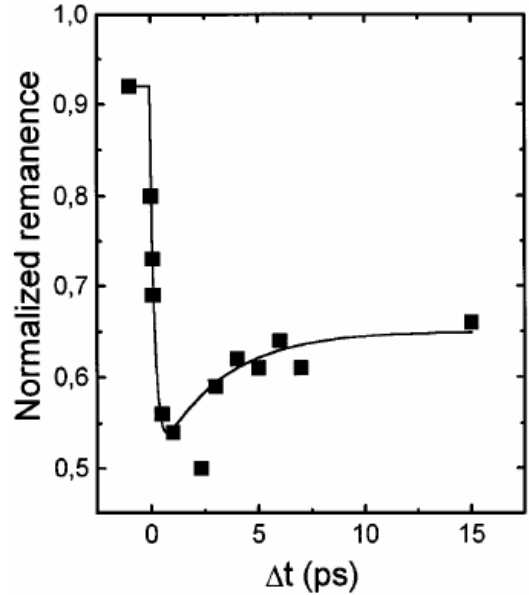,width=8cm,angle=0}}
\caption{ Normalized remanence as a function of time delay between
  pump and probe pulses on a Ni(20 nm)/MgFe$_2$(100 nm) film. The pump
  fluence is 7 mJ/cm$^{2}$. Adopted from Beaurepaire {\it et al.}'s
  original work,\protect\cite{eric} used with permission from the
  American Physical Society.  }
\label{ericfig}
\end{figure}

With the advent of ultrafast laser technology in the 80s and 90s, both
electron and spin dynamics could be detected in the real time domain.
Vaterlaus \et\cite{vaterlaus} reported an interesting observation that
the time scale for establishing thermal equilibrium between the
lattice and spin system is 100$\pm$80 ps in ferromagnetic
gadolinium. This was corroborated theoretically by H\"ubner and
Bennemann.\cite{wolfgang96} A major breakthrough came when Beaurepaire
and his coworkers\cite{eric} found that a 60-fs pulse could
demagnetize the ferromagnetic nickel film within one picosecond (see
Fig. \ref{ericfig}). Such a short reduction was unexpected from the
conventional spin wave theory, and motivated intensive research over
two decades. The research of the first six years has been documented
in our first review paper.\cite{ourreview} Kirilyuk and
coworkers\cite{rasingreview} summarized the research up to 2010. The
reader may refer to these two review papers for those earlier studies.
Kirilyuk and coworkers published another two review papers, one
later\cite{rasingreview2} and one earlier,\cite{rasingreview3} but
there are some overlaps among them. The development of femtomagnetism
also appears in several monographs.\cite{sd1,sd2,sd3,sto,umc13}

Since our review mainly focuses on all-optical spin switching, we
think that it is appropriate to review and give credit to some of
those latest outstanding investigations that have greatly enriched our
understanding of femtomagnetism and directly led to the discovery of
all-optical spin switching.

After the discovery of femtomagnetism by Beaurepaire {\it et
  al.},\cite{eric} almost immediately at least four different
experimental techniques were used.  Scholl \ete\cite{scholl} employed
spin-resolved two-photon photoemission to show that the Stoner
excitation is responsible for the reduction of spin polarization over
1 ps, but unfortunately this result was never reproduced.  Hohlfeld
\ete\cite{hohlfeld1997} employed pump-probe second-harmonic generation
(SHG) with 150-fs/800-nm laser pulses of various fluences and
demonstrated that the transient magnetization reaches its minimum at
50 fs.  Aeschlimann \ete\cite{aes} employed time- and spin-resolved
two photon photoemission to show that the lifetime of majority-spin
electrons at 1 eV above the Fermi energy is twice as long as that of
minority-spin electrons in fcc Co.  La-O-Vorakiat \ete\cite{chan}
developed a tabletop high-harmonic soft x-ray source to probe the
demagnetization at the M edge of Fe and Ni.  Koopmans
\ete\cite{kooprl00} questioned the validity of time-resolved
magneto-optics to detect the spin change.  Three subsequent studies by
Guidoni \ete\cite{guidoni}, Bigot \ete\cite{bigot2004} and Vomir
\ete\cite{vomir} found that time-resolved magneto-optics can be
reliably used, which was further clarified theoretically.\cite{np09}
Regensburger \ete\cite{ren} investigated surface magnetism using
time-resolved SHG. We proposed the first theory of ultrafast
demagnetization\cite{prl00} and showed that the cooperation between
the laser field and spin-orbit coupling is crucial to understanding
the ultrafast demagnetization. Ju \ete\cite{ju} extended research to
exchange coupled systems and found a large modulation in the exchange
bias field.  Kise \ete\cite{kise} found the crossover from microscopic
photoinduced demagnetization to macroscopic critical behavior, with a
universal power law divergence of the relaxation time in half-metallic
Sr$_2$FeMoO$_6$.  Zhang \ete\cite{qzhang} showed that the
magneto-crystalline anisotropy is modulated by nonthermal hot electron
spins in another half-metallic CrO$_2$.  More extensive studies were
carried out by M\"uller \ete\cite{muller} They found that because the
direct energy transfer by Elliot-Yafet scattering is blocked in a
half-metal, the demagnetization time is significantly slower.

 Kimel \ete\cite{kimel2002} discovered antiferromagnetic order
 quenching in FeBO$_3$.  Rhie \ete\cite{rhie} observed exchange
 splitting change in Ni. Gomez-Abal \ete\cite{gom} demonstrated that
 the switching in antiferromagnetic NiO is faster, which was confirmed
 experimentally by Duong \ete\cite{duo}.  Satoh \ete\cite{satoh2010}
 used circularly polarized pulses to drive spin oscillation in the
 fully compensated antiferromagnet NiO.  Wang \ete\cite{wang}
 demonstrated a similar quenching of ferromagnetism in InMnAs and
 enhancement of ferromagnetism in GaMnAs.\cite{wang2007} Stamm
 \ete\cite{stamm} claimed that the quenching of spin angular momentum
 and its transfer to the lattice, with a time constant of 120$\pm$70
 fs, is determined unambiguously with x-ray magnetic circular
 dichroism. Such an unusual fast time scale is surprising. Koopmans
 \ete\cite{koopmans2009} showed that a model based on
 electron-phonon-mediated spin-flip scattering explains both
 timescales on equal footing, but Carva \ete\cite{carva} showed that
 the Elliott-Yafet electron-phonon effect is extremely small.
 Theoretically, Mueller \ete\cite{mueller} suggested a dynamical
 Elliott-Yafet mechanism to dynamically modify the Stoner exchange
 splitting to get the same amount of the spin moment reduction as seen
 in the experiment, and they suggested that in general Elliott-Yafet
 %KEEP spin-flip scattering needs to be taken into account to obtain a
 microscopic picture of demagnetization dynamics is viable.  Very recently,
 Bonetti \ete\cite{bonetti} demonstrated that only the amorphous GdFeB
 sample shows ultrafast demagnetization caused by the spin-lattice
 depolarization of the THz-induced ultrafast spin
 current. Quantitative modeling shows that such spin-lattice
 scattering events occur on similar time scales as conventional
 spin conserving electronic scattering (about 30 fs).  A recent study
 of the phonon angular momentum transfer found a very small effect,
 with the rate of change on the order of 0.06 $\hbar$/100
 fs.\cite{tsa} To diminish the fcc Ni spin moment takes 1
 ps. Thus, the phonon is unlikely the main course of demagnetization
 over a few hundred femtoseconds, as already shown by Lefkidis
 {\it et al.},\cite{lefkidismmm} but there is currently no agreement on this.

Andrade \ete\cite{andrade} studied the response of the magnetization
vector of superparamagnetic nanoparticles and showed that the
magnetization precession is damped more quickly in the superparamagnetic
particles than in cobalt films.  Pickel \ete\cite{pickel} investigated
how the spin-orbit hybridization bands are changed in fcc Co and
demonstrated that these spin hot spots enhance spin-flip scattering.
We showed that the spin and orbital momentum excitation proceed on
different time scales.\cite{prl08} Schmidt \ete\cite{schmidt}
demonstrated that magnon emission by hot electrons occurs on the
femtosecond time scale. Battiato \ete\cite{battiato} proposed
superdiffusive spin transport as a mechanism of ultrafast
demagnetization. Rudolf \ete\cite{rudolf} showed that indeed ultrafast magnetization enhancement in metallic multilayers is
driven by a superdiffusive spin current.  Eschenlohr
\ete\cite{eschenlohr} moved one step further and declared the
ultrafast spin transport as key to femtosecond
demagnetization. Vodungbo \ete\cite{vodungbo} thought that Eschenlohr
\ete's study does not unambiguously prove that part of the magnetization
is transferred to the metallic cap layer via superdiffusive spin
transport. Instead, they claimed that their data undoubtedly prove that
the ultrafast demagnetization process can be triggered without any
direct interaction between the photons of the optical pump pulse and
the magnetic layer. They also emphasized that even though their
results appear in agreement with the superdiffusion prediction, this is not evidence or proof of superdiffusion. A true verification of spin superdiffusion requires a measurement of the spin that is diffused out of the magnetic sample, which was not done in their experiment.\cite{vodungbo} Turgut \ete\cite{turgut} used
multilayers of Fe and Ni with different metals and insulators as the
spacer material to conclusively show that spin currents can have a
significant contribution to optically-induced magnetization dynamics,
in addition to spin-flip scattering processes. Schellekens
\ete\cite{schellekens} found that the temporal evolution of the
magnetization for front-side and back-side pumping in Ni thin films is
identical, so that no influence of transport is detected. Whether
superdiffusion is important is hotly debated for the moment.

Besides those common transition metals, rare earth metals were also
investigated. Lisowski \ete\cite{lisowski} used time-resolved
photoemission and magneto-optics to investigate the spin dynamics on a
Gd(0001) surface, and proposed a mechanism of electron-magnon
interaction to facilitate electron-spin-flip scattering among
spin-mixed surface and bulk states.  Melnikov \ete\cite{melnikov}
looked at magnetic linear dichroism at a core-level photoemission and
showed that equilibration between the lattice and spin subsystems
takes about 80 ps in Gd.  Radu \ete\cite{radu2009} explored the effect
of Ho, Dy, Tb, and Gd impurities on the femtosecond laser-induced
magnetization dynamics of thin Permalloy films using the time-resolved
magneto-optical Kerr effect, and found a gradual change of the
characteristic demagnetization time constant from 60 to 150 fs.  In Gd
and Tb, Wietstruk \ete\cite{wietstruk} observed a two-step
demagnetization with an ultrafast demagnetization time of 750 fs,
identical for both systems, and slower times which differ with 40 ps
for Gd and 8 ps for Tb. Carley \ete\cite{carley} employed the
high-order harmonic generation technique\cite{chan} to investigate the
temporal evolution of the exchange-split bands and found an ultrafast
drop of the exchange splitting. There is a difference in the ultrafast
dynamics between itinerant and localized magnetic moments in
gadolinium metal.\cite{frietsch}

Kim \ete\cite{kim} extended the study to ultrafast magnetoacoustics
and showed that the propagating strain associated with the acoustic
pulses modifies the magnetic anisotropy and induces a precession of
the magnetization.  Theoretically, the  time-dependent density
functional theory was implemented to investigate the spin moment
change.\cite{krieger} Recently, we developed a new technique to
couple the time-dependent Liouville equation with the density
functional theory, where we found that the functional is extremely
important to yield the same amount of the spin reduction.\cite{jpcm16}

To start with, we would like to add a word of caution on the name
conventions used in the all-optical spin switching.  All-optical spin
switching, or AOS, may appear in the literature as all-optical
helicity-dependent spin switching or AO-HDS or AOS. All-optical
helicity-independent spin switching, AO-HIDS, is called ``thermal
switching'' in the literature. Using the word ``thermal'' is
confusing, since a true thermal process should only pertain to the
temperature gradient across a sample from a slow heating source.  Many
research papers do not distinguish the thermal equilibrium among the
electrons or between the electron and phonon subsystems.  In some
studies, the linearly polarized laser pulse is called a heating
pulse. This is also inappropriate since the linearly polarized light
also changes the orbital angular momentum $\Delta l =\pm 1$. The
helicity-independent demagnetization is called thermal demagnetization
(TD).  Conceptually, this is also problematic, since there is a clear
distinction between a true thermal source driven and laser-driven
demagnetization, in particular, on the time scale of a few hundred
femtosecond to 10 ps.  In this paper, we prefer to use AO-HDS and
AO-HIDS. Whenever possible, we will state clearly what the thermal
effect indeed refers to.  Another confusion is that many researchers
also use the time-dependent temperature.  Temperature is a statistical
concept, averaged over a period of time. What is really meant here is the
nonequilibrium population distribution is created in the time domain.

The rest of the paper is arranged as follows. In Section II, we review
the essential experimental findings. Since there are many groups
contributing to the same topics, we group them into five major topics:
(1) Dependence on the sample temperature and compensation temperature;
(2) Dependence on laser parameters; (3) Composition dependence; (4)
Beyond GdFeCo; and (5) Proposed mechanisms. Since some of those topics
slightly overlap, we prefer to repeat the same information in each of
those relevant subsections. Section III is devoted to the
phenomenological theory, which is the most popular theory at this
time. In Section IV, we review our microscopic theory of spin
reversal, starting from the optical spin switching rule, the single
model under cw excitation, spin reversal theory under pulse laser
excitation, inclusion of exchange coupling, demonstration of the
inverse Faraday effect, and finally the effect of the laser field
amplitude on spin reversal. Section V presents some new techniques and
possible new directions. Finally, we conclude this paper in Section
VI.

\section{Experimental findings}

\begin{figure}[th]
\centerline{\psfig{file=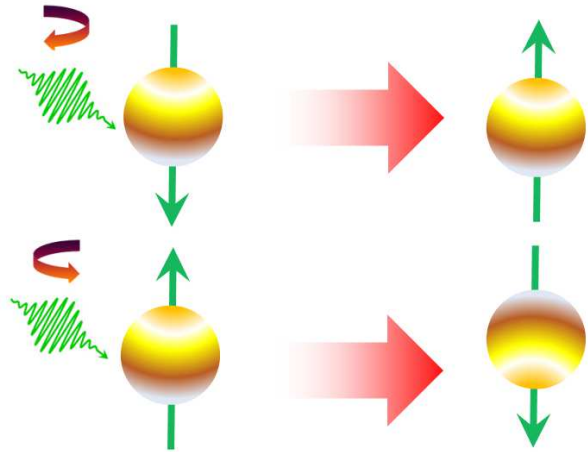,width=11cm,angle=0}}
\caption{
Left-circularly polarized light can switch a
spin from down to up, while right-circularly polarized light can
switch a spin from up to down. A magnetic field is not needed.
 }
\label{fig0}
\end{figure}

Different from the magnetic field-driven spin reversal, AOS relies on
a laser field to switch spin from one direction to another. Figure
\ref{fig0} shows that left-circularly polarized light can switch a
spin from down to up, while right-circularly polarized light can
switch a spin from up to down. This is considered remarkable since the
laser field (E-field) does not directly interact with the spin degree
of freedom.

\begin{figure}[th]
\centerline{\psfig{file=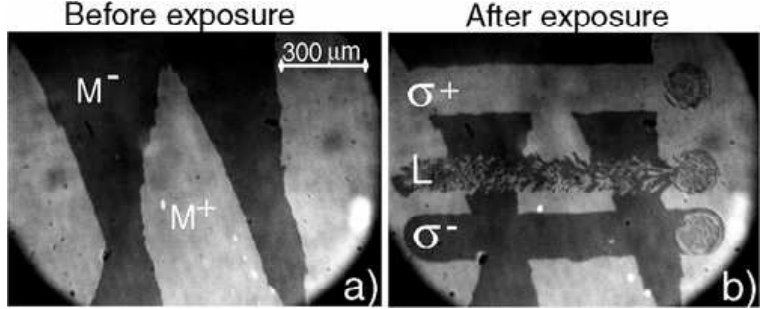,width=10cm}}
%\centerline{\psfig{file=stanciu.fig1.eps},width=5cm}}
\vspace*{8pt}
\caption{ %\cite{stanciu} must be inserted later.
All-optical spin
  switching in \gd{22}{74.6}{3.4} demonstrated by Stanciu
  \ete\protect\cite{stanciu} (a) Magneto-optical image of the initial
  magnetic state of the sample before laser exposure. White and black
  areas correspond to up (M$^+$) and down (M$^{−-}$) magnetic domains,
  respectively. (b) Magnetic domains after sweeping at low speed
  $\mu$m/s linear (L), right-handed ($\sigma^+$), and left-handed
  ($\sigma^-$) circularly polarized beams across the surface of the
  sample, with a laser fluence of about 11.4  mJ/cm$^2$. The
  circularly polarized light switches the magnetization, but the
  linearly polarized light breaks up the domains.  Used with permission
  from the American Physical Society.}
\label{figstanciu}
\end{figure}

In 2007, Stanciu and coworkers\cite{stanciu} reported that a single
40-fs laser pulse, free of any magnetic field, could switch the spin
in \gd{22}{74.6}{3.4} from one direction to another
(Fig. \ref{figstanciu}). If left-circularly polarized light (LC)
switches the spin from down to up, then right-circularly polarized
light (RC) switches the spin from up to down. Linearly polarized light
(LP) induces small domains with spin randomly oriented up or
down. This is the beginning of AO-HDS.  Different from regular
ferromagnetic thin films, GdFeCo is very complex.  Structurally,
metallic \gd{22}{74.6}{3.4} is amorphous, posing a big challenge for a
direct simulation which requires the structural information.
Optically, the material is highly absorptive.  Magnetically, two
magnetic sublattices are coupled ferromagnetically, i. e., the
coupling between Gd ions and that between Fe ions are
ferromagnetic. But the coupling between Gd and Fe is ferrimagnetic,
with the exchange interaction on the order of meV.\cite{man,taylor}
Gd$_{22}$Fe$_{74.6}$Co$_{3.4}$ has a saturation magnetization of 1000
G at room temperature and Curie temperature of 500 K.

AOS depends on the sample's temperature, compensation temperature,
composition, and laser parameters.  To understand the microscopic
mechanism of how AOS works, one has to disentangle the convoluted effects
both intrinsically and extrinsically. In the following, we make a
moderate attempt to present crucial experimental facts, even when
they sometimes are contradictory among themselves.

\subsection{Dependence on sample and compensation temperature}

One unique feature of ferrimagnets is that the system has two magnetic
sublattices, $A$ and $B$. Here $A$ refers to Gd and $B$ refers to
Fe/Co.  The spin moments on $A$ and $B$ are different and in the
opposite direction. We define the compensation temperature when the
spin moments from sublattices are equal in magnitude but point in
the opposite direction. Thus they cancel each other out, and the net
magnetization of the entire sample drops to zero.

Vahaplar \ete\cite{vahaplarprb} found that the optimal conditions for
the all-optical reversal are achieved just below the ferrimagnetic
compensation temperature.  The sample temperature also influences how
large the laser fluence should be in order to switch
spins.\cite{vahaplarprb} The larger the deviation of the sample
temperature from the compensation temperature, the higher the laser
fluence required for the switching.\cite{vahaplarprb} Vahaplar \et
suggested that choosing temperature in the vicinity of the
compensation temperature is very important for the spin reversal.
Hohlfeld \ete\cite{hohlfeld} showed that too high a temperature is
detrimental to AOS.  In an earlier study,\cite{vahaplarprl} Vahaplar
\et found that the spin reversal time in
Gd$_{24}$Fe$_{66.5}$Co$_{9.5}$ is on the order of 90 ps, and does not
change much among GdFeCo alloys if the sample temperature is below the
compensation temperature. But once the temperature is above the
compensation temperature, the time increases sharply.

However, the above finding is not generic across all AOS materials. In
2013 Hassdenteufel \ete\cite{hass2013} found that AOS in
\tbfe{x}{100-x} occurs below and above the magnetic compensation
point, and they even found that AOS takes place in samples without a
compensation temperature.

\subsection{Dependence on laser parameters}

\begin{figure}[th]
\centerline{\psfig{file=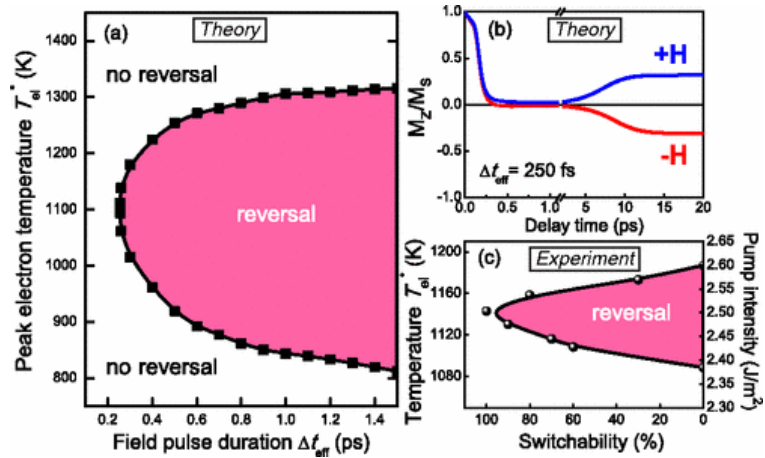,width=10cm}}
%\centerline{\psfig{file=stanciu.fig1.eps},width=5cm}}
\vspace*{8pt}
\caption{ Spin reversal only occurs in  a narrow region. (a) Theoretical results from the phenomenological simulation. (b) Magnetic field-driven switching. (c) Experimental results. Note that the shape is skinny, different from the theory.\protect\cite{vahaplarprl} Used with permission from the American Physical Society.}
\label{vahaplarfig}
\end{figure}

The dependence of the laser fluence on the AOS was first explored.
Stanciu \ete\cite{stanciu} showed that at a laser fluence of 2.9
mJ/cm$^2$, only one type of helicity, LC or RC, can reverse the spin.
When they increased the fluence to 5.7 mJ/cm$^2$, regardless of the
laser helicity, multiple domains were formed after the laser exposure.
Vahaplar \ete\cite{vahaplarprl} found that AOS in \gd{22}{68.2}{9.8}
only occurs in a narrow fluence range, and the switchability forms a
``$\Lambda$'' shape (see Fig. \ref{vahaplarfig}).\footnote{We note in
  passing that their fluence has a unit of J/m$^2$, but in their
  latter paper,\cite{vahaplarprb} they changed it back to mJ/cm$^2$,
  so it is difficult to know at present which unit they referred to.}
In \gd{24}{66.5}{9.5} and \gd{26}{64.7}{9.3} the reversal window gets wider as the laser pulse duration increases.\cite{vahaplarprb}
Chen \ete\cite{chen} directly measured the hysteresis loop in
\gd{23.5}{73.2}{3.3} as a function of the external magnetic field with and
without the pump pulse. Interestingly, they found that the hysteresis
becomes anomalous, consistent with Stanciu \ete, \cite{stanciu2} and is no longer a
square shape. Instead, the domain breaks into different parts, some of
which become irreversible. This is reflected on the hysteresis
loop. Stanciu \ete\cite{stanciu2} also performed a laser fluence-dependent study of the hysteresis and found that when the pump fluence
is higher, the spin first relaxes in the opposite direction; then
once the system cools down, the spin reverses back to the initial
configuration.

A similar fluence dependence in \gd{26}{65}{9} was carried out by
Khorsand \ete\cite{khorsand} who showed that the switching probability
increases with the fluence, but differently for left-circularly
polarized light, right-circularly polarized light and linear polarized
light. They explained the difference by different optical absorption
efficiencies among different helicities. The effective switching
threshold is independent of the wavelength, at $2.6\pm 0.2$ mJ/cm$^2$,
which is lower than \gd{26}{64.7}{9.3}.\cite{vahaplarprb}

A more systematic investigation of the effect of the laser parameters,
including, the laser pulse duration, wavelength, chirp and bandwidth,
was performed by Steil \ete\cite{steil} They showed that AOS in
\gd{26}{64.7}{9.3} can be achieved with a picosecond laser as
well, and the microscopic process seems only to depend on the number
of photons.

\begin{figure}[th]
\centerline{\psfig{file=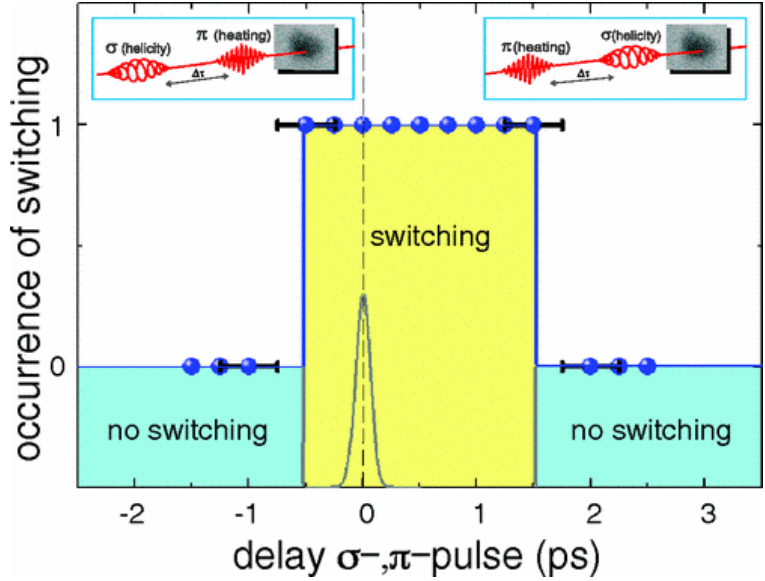,width=10cm}}
%\centerline{\psfig{file=stanciu.fig1.eps},width=5cm}}
\vspace*{8pt}
\caption{ Occurrence of switching as a function of time delay between
  $\sigma$ and $\pi$ pulses. Only a limited switching window is
  observed.\protect\cite{sabine2012} Used with permission from the American
  Physical Society.}
\label{sabinefig}
\end{figure}

Alebrand \ete\cite{sabine2012} provided a much needed insight. They
employed different combinations of two laser pulses, one linearly
polarized light ($\pi$) and the other circularly polarized light
($\sigma$) (see Fig. \ref{sabinefig}).
Their goal was to distinguish
the effects of heating and helicity of the laser pulse.  Their idea is
that the $\sigma$ pulse carries the helicity information while the
$\pi$ pulse provides heating. Using different combinations, one may be
able to differentiate the roles of the laser pulse from others. Before
we proceed, we note that both $\sigma$ and $\pi$ carry the helicity
information and heating.  Their sample was \gd{24}{66.5}{9.5}, with
out-of-plane magnetization, compensation temperature of 280 K and
Curie temperature of 500 K. The central wavelength of their laser is
780 nm, with pulse duration changeable from 90 fs to 2 ps. They first
used a single $\sigma$ pulse and lowered its fluence until no
switching is observed, and then they added a $\pi$ pulse. They found
that the switching becomes possible again as far as these two pulses overlap
spatially and temporally.  If they increased the $\pi$
pulse fluence further, a helicity-independent demagnetization was
observed.The total threshold fluence for the $\sigma-\pi$ combination
is always higher than a single $\sigma$ pulse minimum threshold
fluence. This indicated that the switching does not only depend on the
number of photons, different from their original finding,\cite{steil}
but also on the helicity. The circularly polarized light appeared more
powerful. They also decreased the $\sigma$ pulse fluence and found
that there exists a smaller threshold for the $\sigma$ pulse once the
$\pi$ pulse is on. However, once the fluence for the $\sigma$ pulse is
below the above lower threshold, increasing the $\pi$ fluence can not
lead to switching. Instead, a pure demagnetization (helicity
independent) occurs.

Alebrand \ete \cite{sabine2012} also investigated how the minimum
threshold fluence of the circularly polarized light depends on the
laser repetition rate, and they found that as the repetition rate
increases from 0 Hz to 500 kHz, the required minimum fluence drops
from 6 to 1.5 mJ/cm$^2$. It is unclear whether this finding
is related to a recent study by El Hadri \ete\cite{hadri} who employed
the Hall cross to characterize the switching,\cite{elhadri} and found
that in GdFeCo the switching is ``single pulse'' instead of
``cumulative.''

\subsection{Composition dependence}

The investigation of the composition effect on AOS is much more
extensive than any other studies because AOS does not happen in any
composition.  This is easy to see this from their element spin moments.
Gd has nearly zero orbital angular momentum since its 4f orbital is
half-filled, but the spin moment is big, and in pure Gd metal it is
7.63 $\mu_B$.\cite{kurz,jensen} By contrast, iron has a spin moment of
2.2 $\mu_B$.  The small composition of Co is to control the
perpendicular anisotropy.\cite{ostler}

Vahaplar \ete\cite{vahaplarprb} employed five different compositions:
\gd{20}{70}{10}, \gd{22}{68.2}{9.8}, \gd{24}{66.5}{9.5},
\gd{26}{64.7}{9.3}, and \gd{28}{63}{9}.  With the same laser fluence,
duration and polarization, they showed that different compositions of
Gd ions have a significant effect on AOS. At 3.14 mJ/cm$^2$, only
\gd{26}{64.7}{9.3} shows helicity-dependent switching, while
\gd{22}{68.2}{9.8} and \gd{24}{66.5}{9.5} show
helicity-independent switching.  But by lowering the pump fluence,
\gd{22}{68.2}{9.8} and \gd{24}{66.5}{9.5} also show
helicity-dependent switching. However, \gd{20}{70}{10} is very
different from the rest of the compositions. A single laser of any
polarization only leads to a multidomain state. This clearly
demonstrates the decisive role of the composition in AOS. Future
research should focus on much more on this interesting development.

\begin{figure}[th]
\centerline{\psfig{file=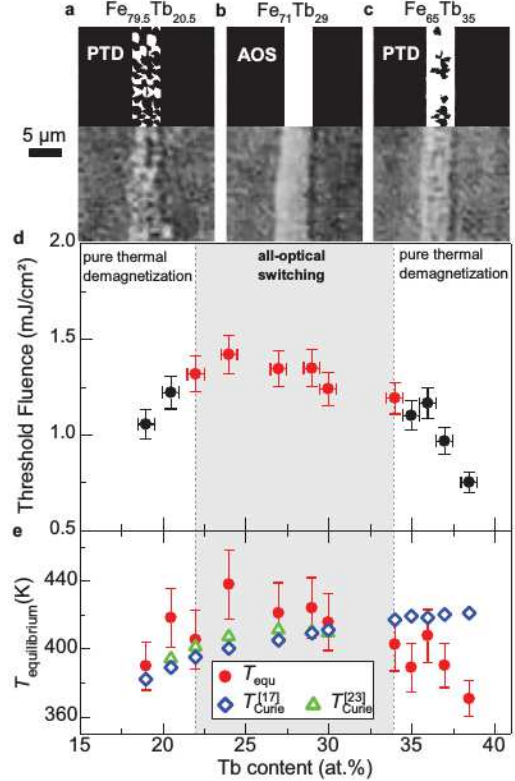,width=7cm}}
%\centerline{\psfig{file=stanciu.fig1.eps},width=5cm}}
\vspace*{8pt}
\caption{ In  \tbfe{x}{100-x},  AOS occurs only between $x=22\%$ and
  $x=34\%$.\protect\cite{hass2013} Used with permission
  from Advanced Materials.}
\label{hassfig}
\end{figure}

The composition effect is also observed in \tb{x}{1-x} alloys.
Alebrand \ete\cite{sabineapl} showed that the optical magnetization
switching is observed only for $x=26\%, 28\%$ and 30\%.  It is in this
composition region where the compensation temperature is higher than
the room temperature but lower than the Curie temperature. From this
study, it is clear that composition plays a decisive role here too,
but it is unclear whether the compensation temperature is a result of
composition or the cause of AOS.  In a latter study,\cite{sabine2014}
they reported that in \tb{32}{68} a transient magnetization reversal
occurs on both sublattices on a subpicosecond time scale, but AOS was
not observed in this compound. This may indicate that a transient
reversal is not a necessary precursor to spin reversal, as it is in
GdFeCo\cite{radu} and TbFe\cite{khorsandprl} where a transient
ferromagnetic-like state is identified before spin reversal, but their
technique is not really element-specific, and their entire observation
relied on the wavelength dependence. In \tb{26}{74} where AOS was
observed, a transient reversal was observed again, but only at a probe
wavelength of 800 nm and at a high laser fluence of 2.4 mJ/cm$^2$. In
summary, since their technique is not really element-specific, it is
difficult to determine whether the actual spin reversal occurred (or
not) at which site Co or Tb.

In 2013, Hassdenteufel \ete\cite{hass2013} systematically investigated
the role of the composition on the AOS in \tbfe{x}{100-x} and found
that AOS occurs only between $x=22\%$ and $x=34\%$. Outside this
composition range, only helicity-independent demagnetization is
observed (see Fig. \ref{hassfig}).

 Mangin \ete\cite{mangin} summarized the trend for AOS in different
 compounds. In GdFeCo, the Gd composition must be between 25\% and 30\%;
 in TbCo, the Tb composition between 22\% and 26\%; in DyCo with Dy
 composition between 26\% and 29\%; in HoFeCo with Ho composition
 between 23\% and 26\%; in TbCo multilayers, 17\%-32\%; and in HoCo, the
 Ho composition must be between 22\% and 31\%. The understanding of
 these distinctive composition dependencies is lacking.

\subsection{Beyond GdFeCo}

Most AOS materials are GdFeCo and its variants with different
compositions. As discussed above, Alebrand \ete\cite{sabine2012}
discovered spin switching in TbCo alloys, while Hassdenteufel
\ete\cite{hass2013} showed TbFe works as well. In 2014, Mangin
\ete\cite{mangin} discovered that the rare-earth-free Co-Ir-based
synthetic ferrimagnetic heterostructures are also AOS materials. AOS
is found in Co(0.5nm)/Tb(0.4nm) multilayers, but not
Co(0.8nm)/Tb(0.4nm). Note that AOS only favors ultrathin Co, a
peculiar feature that is not yet understood.  Lambert and
coworkers\cite{lambert} showed that AOS occurs in ultrathin
ferromagnetic CoPt films. The number of repeats is between 2 and 3; if
higher than this, only helicity-independent demagnetization is
found. The thickness of the Co layer must be below 1.0 nm; and the
laser power threshold is below 500 nW. By increasing the Ni
concentration, one gets thermal demagnetization, not AOS. These
stringent conditions are interesting and useful for future research.
There is some similarity between different AOS materials.  Some
simulations on ferromagnets have appeared,\cite{corn,bokor2016} but a
complete picture is missing. In particular, there has been very few
independent studies from other groups.

\subsection{Proposed mechanisms}

So far, there are several mechanisms proposed. According to Kirilyuk
\ete\cite{rasingreview} the interactions between the laser and magnetic
systems fall into three categories: (a) thermal effects, (b)
nonthermal photomagnetic effects, and (c) nonthermal optomagnetic
effects.  Optomagnetism refers to a magnetic process where the system
does not directly absorb the light energy, but may interact with other
elementary excitations such as phonons or magnons. Photomagnetism
refers to a magnetic process where the system absorbs energy
from the light field.

Stanciu \ete\cite{stanciu} suggested stimulated Raman scattering for
AOS that involves angular momentum transfer from the lattice to
the spins. Their suggestion was based on a simple estimate that
switching due to direct transfer of angular momentum via absorption is
unlikely.  However, Hohlfeld \ete\cite{hohlfeld} looked at the temperature
dependence of AOS on the same metallic sample as Stanciu \ete\cite{stanciu}  Hohlfeld \et found that at 2.5 mJ/cm$^2$, which is
below the laser fluence threshold of about 3 mJ/cm$^2$, the switching
is not possible between 250 and 300 K, but as the temperature is
lowered, switching occurs. They concluded that increasing the
temperature reduced the efficiency of the AOS, which invalidated the
original ideas based on stimulated Raman-like
scattering.\cite{stanciu}  This result is consistent with the study
with different substrates.\cite{hass2014} Hassdenteufel \et showed
that for \tbfe{30}{70}, AOS occurs on the SiO$_2$/Si substrate but not
on the microscope glass slide.\cite{hass2014} It seems that heating
accumulation with a higher temperature is detrimental to AOS.

Vahaplar \ete\cite{vahaplarprl} suggested the inverse Faraday effect
as the driving field to AOS.  In 2012, Ostler and
coworkers\cite{ostler} found that when the laser fluence is above a
certain threshold value at which Vahaplar \ete\cite{vahaplarprl} found
helicity-dependent switching, the switching is helicity-independent.
Ostler \et claimed that their new observation negates their own
explanation based on the inverse-Faraday effect.  They called this new
AOS a thermal effect (i.e., helicity-independent switching).  This
result reflects the complex nature of AOS in GdFeCo.
To clarify the situation, we notice that, according to Ostler {\it et
  al.}, to go from helicity-dependent to helicity-independent spin
switching, one only needs to increase the laser fluence by as little
as 0.05 mJ/cm$^2$. This is rather surprising.  We estimate that for a
0.05 mJ/cm$^2$ change in fluence, the photon number increases by 0.05
photon.\cite{jpcm13} It is unclear as to how reliable their
characterization is. This problem remains unsolved.

Steil \et showed that while the
concept of an induced effective magnetic field $H_{eff}$ agreed with
their experimental results,\cite{steil}  the field could not be
explained in terms of an inverse Faraday effect nor spin-flip
stimulated Raman scattering.  They also wondered where the helicity
information is stored after the duration of the laser pulse. A few
years earlier, Stanciu \ete\cite{stanciu} already argued that the
actual magnetization reversal must take place on a subpicosecond time
scale, since the interaction between the laser field and the sample
lasts only 40 fs (pulse duration).  This is the only way for the
helicity information to survive on a long time scale. Otherwise, why
doesn't the lattice mess up the angular momentum?

Khorsand \ete\cite{khorsand} proposed magnetic circularly dichroism
(MCD) as the underlying mechanism for AOS. They claimed that they
could quantitatively explain AOS using MCD.  What they did is to
compare the inverse Faraday effective magnetic field with magnetic
circularly dichroism or magnetic circular birefringence, and to see
which one has a larger effect. There is no actual time-dependent
simulation to prove that MCD is responsible for AOS.  Therefore, it
may be too early to claim quantitative proof as it intrinsically
excludes other possible mechanisms.\cite{berritta} It may be more
appropriate to consider all the possible scenarios before a definitive
statement can be made. For instance, they initially insisted that AOS
is an optomagnetic process,\cite{vahaplarprb} but now they favor the
photomagnetic process, where the AOS depends on the amount of energy
absorbed from the laser.\cite{khorsand}

In searching for the essence of AOS, Schubert \ete\cite{schubert}
proposed that the low remanent magnetization is a key prerequisite for
AOS. They chose \tbfe{36}{64} and \tbfe{19}{81}, none of which shows
AOS. But when they exchange coupled them to form a heterostructure
with zero spin moment, then AOS occurs in the combination. This
demonstrates the importance of remanence on AOS. This low-remanence
criterion was further reinforced by Hassdenteufel
\ete\cite{hass2015} They found that a low remanent sample
magnetization is crucial for all-optical magnetic switching in
ferrimagnets and ferrimagnet heterostructures. For nearly all the AOS
materials, the remanence is below 220 emu/cc. The small remanence
criterion, if verified by other groups, is extremely interesting.  In
2013, Barker \ete\cite{barker} suggested that a two-magnon bound state
causes ultrafast thermally-induced magnetization switching.

\section{Phenomenological theory}

Since the discovery of AOS by Stanciu \ete,\cite{stanciu} several
theoretical approaches have appeared. However, the majority are
phenomenological and do not have actual calculations.
Gridnev\cite{gr} developed a theory for a transparent magnetic
dielectric, where the key interaction is impulsive stimulated
Raman scattering. The time dependence of the effective field is
approximated by a $\delta$-function. However, in the real experiment,\cite{stanciu} GdFeCo is metallic and not transparent, so that a direct
application of the theory to those materials is difficult. Along the
same line, Popova \ete\cite{pop} adopted a hydrogenlike system. They
employed the same technique used by Lefkidis \ete\cite{george} and
were able to factor out the induced magnetization. The insight they revealed is that in contrast to the previous belief in the
inverse Faraday effect, where the effective magnetic field is
proportional to ${\bf E}(t)^*\times {\bf E}(t)$, the actual effective
magnetic field is much more complicated. Only in the limit of a very
long pulse or cw wave is the above relation restored, which is
already clear from the original work by Pershan \ete\cite{pershan} The
same conclusion is found under the Drude-Lorentz
approximation.\cite{battiato2014} Unfortunately, ${\bf E}(t)^*\times
{\bf E}(t)$ is still in use up to now. In 2013, Gridnev\cite{gr13}
proposed an interesting approach where the itinerant electrons are
heated by the laser. The spin polarization is changed through the rate
equation but augmented by a spin generating function $G$. $G$ is
proportional to the laser intensity multiplied by a conductivity
tensor and energy conservation $\delta$-function. This approach is
reasonable if under cw excitation, but when using a pulsed laser, this
is not appropriate, since $G$ changes as the laser field. Therefore,
in the real calculation, the author replaced $G$ by a special initial
value of the spin, which makes the theory highly empirical.

Ostler \ete\cite{ostler2011} employed fcc cells under periodic
boundary conditions, where the transition metal (TM) and rare-earth
(RE) ions are randomly distributed. Such a procedure does not take
into the distance between ions. The entire system is described by the
following Hamiltonian: \be H=-\frac{1}{2}\sum_{ij} J_{ij} {\bf
  S}_{i}\cdot {\bf S}_j -\sum_i D_i({\bf S}_i\cdot {\bf n}_i)^2
-\sum_i \mu_i {\bf B}\cdot {\bf S}_i. \ee Here, $J_{ij}$ is the
exchange integral between spins at site $i$ and $j$, ${\bf S}_i$ is
the normalized vector $|{\bf S}_i|=1$, $D_i$ is the uniaxial
anisotropy vector, $\mu_i$ is the magnetic moment of the site $i$, and ${\bf B}$ is the vector describing the applied field. Their exchange
integrals are $J_{\rm TM-TM}=0.0281$ eV between TMs,
$J_{\rm RE-RE}=0.00787$ eV between REs, and $J_{\rm TM-RE}=-0.0068$ eV
between TM and RE. These values are somewhat changed in their later
studies.\cite{ostler}  It is clear that such a Hamiltonian can not
describe the AOS, since there is no laser field. To overcome this
difficulty, Vahaplar \ete\cite{vahaplarprb} introduced an optomagnetic
field ${\bf H}_{\rm OM}$ by considering the fact that a circularly
polarized subpicosecond laser pulse can act on spins as an effective
light-induced magnetic field. Specifically they used a
phenomenological expression from the inverse Faraday effect derived
for a transparent medium in thermodynamic equilibrium under cw
excitation, \be {\bf H}_{\rm OM}(t,r)=\epsilon_0 \beta [{\bf
    E}(t,r)\times {\bf E}^*(t,r)] \label{hom},\ee where $\epsilon_0$ is
the permittivity in vacuum, $\beta$ is the magneto-optical
susceptibility, and $|{\bf E}(t,r)|$ is the envelope of the laser
E-field. Some comments are necessary. First, Eq. (\ref{hom}) is
conceptually very simple, but the cross product of two electric fields
${\bf E}$ carries different spatial indices,\cite{belotelov} which
are coupled with $\beta$; otherwise, the cross product between two
identical vectors is zero. For this reason, they had to introduce
another coefficient $\sigma$, which is $\pm 1$ for left- and right-circularly polarized light and is 0 for linearly polarized
light. In other words, linearly polarized light has a zero ${\bf
  H}_{\rm OM}$, at variance with the experimental
findings.\cite{stanciu,ostler} There is some inconsistency here.
Second, the original expression is obtained under cw excitation, but
here a pulse laser is used.  If one uses ${\bf H}_{\rm OM}$, then the
entire effective field would have the same duration as the laser
field, which is 40 fs in their case. To overcome this issue, they
split their ${\bf H}_{\rm OM}$ into two half-pulses. The first half
follows the laser field, and the second half does not.  Under the
above approximation, they solved the Landau-Lifshitz-Gilbert equation
to compute the spin reversal.

An extension to the above study was made by Wienholdt {\it et
  al.},\cite{wienholdt} who separated the spins according to their
orbital characters, such as $d$ and $f$.  The entire simulation is
similar to Ostler \ete\cite{ostler} The laser effect is simulated by a
temperature increase, which is normal for this kind of
simulation. They also found transient ferromagnetic ordering.

Another approach is based on the multisublattice magnets.  Mentink
\ete\cite{mentink} coupled the multisublattice to a heat bath with a
time-dependent temperature. The laser field is ignored, and instead is
replaced by a time-dependent temperature. Since temperature is a
statistical concept, introducing a time-dependent temperature is
questionable, but a phenomenological theory was useful in the
beginning of AOS investigation.

Assuming an inverse Faraday
effect, Petrila \ete\cite{petrila} investigated the dependence of AOS on the laser
parameters.  A similar approach was employed by Cornelissen
\ete,\cite{corn} who developed another model simulation to address
switching in the Co/Pt system. This was an interesting approach as it does
not involve some complicated calculation.

 Chimata \ete\cite{chimata} carried out an investigation on a
 supercell with 200 Gd and Fe atoms with amorphous structures at the
 first-principles level, but only at the static structure and magnetic
 properties level. This represents an important improvement over the
 previous studies. However, for spin switching, they still used
 the Landau-Lifshitz Gilbert equation, so there is no laser pulse in
 the simulation; instead they used an effective magnetic field and
 electron temperature. As a result, only the thermal switching was
 investigated. They found the thermal switching was observed for all
 the cases, regardless of whether the initial temperature was above or
 below a compensation temperature.

  Baral and Schneider\cite{bar}
adopted a model that couples the local spins with itinerant spins
antiferromagnetically. They showed that with a sufficiently strong laser, a transient ferromagnetic-like
state can always appear, but this state only results in true spin switching when the
model parameters yield the compensation point.

\section{A simple all-optical spin switching theory}

To this end, it is very difficult to develop a comprehensive theory
for all-optical spin switching. AOS materials are very diverse and
very complex; some of them are amorphous. From the above discussion,
we have seen nearly all theoretical investigations are
phenomenological and mostly build upon an effective magnetic
field. But the interest in AOS is a magnetic field-free spin reversal.
Encouraged by the recent finding of AOS in ferromagnet
CoPt,\cite{lambert} we find a simple way to understand AOS. Most of
the materials here are unpublished and first presented in this
review. The initial finding is very attractive. For this reason, we
provide a MatLab code for our theory, so the reader, in particular,
the experimentalist, can directly adopt it to explain their
experimental results, though our MatLab code\footnote{This
  code will be published in our book entitled {\it Introduction to
    Ultrafast Phenomena: From Femtosecond Spin Dynamics to Attosecond
    High Harmonic generation} by G. P. Zhang, W. H\"ubner, G. Lefkidis,
  A. Rubio and T.  F. George (CRC, Boca Raton, FL, 2018).}does not include the
exchange interaction for the moment.

\subsection{Optical spin switching rule among spin-orbit coupled states}

All-optical spin switching is an optical process, so it must obey the
dipole selection rule.\cite{prb09} But common selection rules are
often restricted to pure spin states, so the spin is unchanged.  We
choose two sets of spin-orbit coupled states, where the total angular
momentum quantum number $j$ and the magnetic one $m_j$ are good
quantum numbers. One may consider these states as a basis of the
eigenstates for a solid, and they take a significant weight on the
true wavefunctions. {Materials of different kinds may have
  different weights on those states and lead to either
  demagnetization, magnetization or spin reversal, or any combination
  of them.}  To develop an analytic expression for the spin switching
is difficult if we directly adopt the eigenstates of solids, partly
because the true physics of AOS is mostly hidden in the lengthy
summation over band states.

In the following, we consider two spin-orbit coupled states, $\psi_a$
and $\psi_b$,\cite{prb09}
\noindent
\ba \psi_a&=\roof{l+m+1}{2l+1}
Y_{lm}|\alpha\ra+\roof{l-m}{2l+1} Y_{lm+1} |\beta\ra,& \label{eq1}
{~~\rm for~~} j=l+1/2, m_j=m+1/2,
\\ \psi_b&=-\roof{l-m}{2l+1} Y_{lm}|\alpha\ra +\roof{l+m+1}{2l+1}
Y_{lm+1} |\beta\ra,& \label{eq2} {\rm ~~for~~} j=l-1/2,
m_j=m+1/2,  \ea where $|\alpha\ra$ and $|\beta\ra$ refer to the
spin-up and spin-down eigenstates, and $Y_{lm}$ is spherical harmonic
with angular and magnetic angular quantum number $l$ and $m$,
respectively.  $l$ and $m$ in Eqs. (\ref{eq1}) and (\ref{eq2}) may
differ from each other.  We should point out that the transition
between $\psi_a$ states or between $\psi_b$ states changes the total
angular momentum quantum number by 1 or $\Delta j=\pm 1$, while the
transition between $\psi_a$ and $\psi_b$ does not change $j$, or
$\Delta j=0$.  The spin angular momenta for the above two states are\cite{prb09}
\ba S_z^a&=&\frac{m_j}{j}\frac{\hbar}{2}\\ S_z^b
&=&-\frac{m_j}{j+1}\frac{\hbar}{2},  \ea
where we see that $\psi_a$ is mainly in a spin-up state while $\psi_b$
is in a spin-down state, if $m_j>0$ is assumed.

To quantify the spin reversal, we employ the dimensionless spin
switchability \be \eta = \frac{S_z^f}{S_z^i}, \ee where $S_z^{i(f)}$
is the initial (final) spin. $\eta >1 $ means that the spin increases
in the original direction of the initial spin; $1>\eta\ge 0$
corresponds to the demagnetization; $-1\le \eta<0$ signifies the spin
reversal; and $\eta < -1$ indicates that the spin is reversed and
increases in the opposite direction of the original spin. Among the
linearly and circularly polarized lights, there are 18 possible spin
changes. They cover the full spectrum of the spin excitation.

%\begin{table}
%%\small
%\caption{Spin switchability $\eta$ of all-optical spin reversal among
%  spin-orbit coupled states. For all $\Delta j=-1$ cases, $j$ must be
%  no less than 3/2.
% LP refers to the linearly
%  polarized light, LC is the left-circularly polarized light and RC
%  the right-circularly polarized light.  }
%\begin{tabular}{l|cc|c|c|cc}\hline\hline
% & \multicolumn{2}{c|}{$\eta^{a\rightarrow a'}$} & {$\eta^{a\rightarrow b}$} & {$\eta^{b\rightarrow a}$}
%& \multicolumn{2}{c} {$\eta^{b\rightarrow b'}$}
% \\
%& $\Delta j=+1$ &  $\Delta j=-1$ & $\Delta j=0$ & $\Delta j=0$  &
%$\Delta j=+1$ & $\Delta j=-1$ \\
%\hline
%LP   & $\frac{j}{j+1}$ & $\frac{j}{j-1}$ &$-\frac{j}{j+1}$ &$-\frac{j+1}{j}$ &$\frac{j+1}{j+2}$ &$\frac{j+1}{j}$\\
%($\Delta m_j=0$)  & & & & & & \\
%\hline
%LC  & $\frac{j(m_j+1)}{(j+1)m_j}$ &$\frac{j(m_j+1)}{(j-1)m_j}$& $-\frac{j(m_j+1)}{(j+1)m_j}$ & $-\frac{(j+1)(m_j+1)}{jm_j}$ &
%$\frac{(j+1)(m_j+1)}{(j+2)m_j}$ & $ \frac{(j+1)(m_j+1)}{jm_j} $\\
%($\Delta m_j=+1$)  & & & & & & \\
%RC  & $\frac{j(m_j-1)}{(j+1)m_j}$
%&$\frac{j(m_j-1)}{(j-1)m_j}$& $-\frac{j(m_j-1)}{(j+1)m_j}$ &
%$-\frac{(j+1)(m_j-1)}{jm_j}$  &
%$\frac{(j+1)(m_j-1)}{(j+2)m_j}$ & $\frac{(j+1)(m_j-1)}{jm_j}$\\
%($\Delta m_j=-1$)  & & & & & & \\
%\hline
%\hline
%\end{tabular}
%\label{table1}
%\end{table}

\begin{table}
%\small
\caption{Spin switchability $\eta$ of all-optical spin reversal among
  spin-orbit coupled states. For all $\Delta j=-1$ cases, $j$ must be
  no less than 3/2.
 LP refers to linearly
  polarized light, LC left-circularly polarized light, and RC
  right-circularly polarized light.  }
\begin{tabular}{lrrr}
\hline\hline
& LP ($\Delta m_j=0$) & LC  ($\Delta m_j=+1$)& RC($\Delta m_j=-1$) \\
\hline
%                                      & ($\Delta m_j=0$) & ($\Delta m_j=+1$) & ($\Delta m_j=-1$)\\
$\eta^{a\rightarrow a'}$  $\Delta j=+1$ &  $\frac{j}{j+1}$
  & $\frac{j}{j+1}\frac{m_j+1}{m_j}$ & $\frac{j}{j+1}\frac{m_j-1}{m_j}$  \\
$\eta^{a\rightarrow a'}$  $\Delta j=-1$ &   $\frac{j}{j-1}$
  & $\frac{j}{j-1}\frac{m_j+1}{m_j}$&$\frac{j}{j-1}\frac{m_j-1}{m_j}$ \\ \hline
$\eta^{a\rightarrow b}$ $\Delta j=0$  & $-\frac{j}{j+1}$ &
  $-\frac{j}{j+1}\frac{m_j+1}{m_j}$ &
  $-\frac{j}{j+1}\frac{m_j-1}{m_j}$\\ \hline
$\eta^{b\rightarrow a}$ $\Delta j=0$  & $-\frac{j+1}{j}$ &
  $-\frac{j+1}{j}\frac{m_j+1}{m_j}$ &
  $-\frac{j+1}{j}\frac{m_j-1}{m_j}$\\ \hline
$\eta^{b\rightarrow b'}$  $\Delta j=+1$ &  $\frac{j+1}{j+2}$
  & $\frac{j+1}{j+2}\frac{m_j+1}{m_j}$ & $\frac{j+1}{j+2}\frac{m_j-1}{m_j}$  \\
$\eta^{b\rightarrow b'}$  $\Delta j=-1$ &   $\frac{j+1}{j}$
  & $\frac{j+1}{j}\frac{m_j+1}{m_j}$&$\large \frac{j+1}{j}\frac{m_j-1}{m_j}$
                               \\
\hline
\hline
\end{tabular}
\label{table1}
\end{table}

We start with linearly polarized light (LP), where $\Delta
m_j=0$. Table \ref{table1} shows that for $\Delta j=+1$, regardless of
whether the transition is between $\psi_a$ states or between $\psi_b$
states, $\eta$ is positive and less than 1, or demagnetization. By
contrast, for $\Delta j=-1$, it corresponds to magnetization
enhancement since $\eta>1$.  $\psi^a$ and $\psi^b$ each have one
demagnetization ($\Delta j=+1$) and one magnetization channel ($\Delta
j=-1$) if the transition is only between $\psi_a$ states or $\psi_b$
states.  These two channels can not be categorized as a thermal
process, since the photon angular momentum is transferred to ($\Delta
j=1$) and away from ($\Delta j=-1$) the system. Quantitatively, $\eta$
depends on $j$. For example, consider $\psi_a \rightarrow \psi_{a'}$,
and with $\Delta j=1$ (demagnetization), if $j=3/2$, $\eta=3/5$, the
percentage spin loss is $1-\eta=2/5$, or 40\%. This means that for a
single photon absorbed, the spin can be reduced by 40\%, which is
compatible to the experimental findings.\cite{eric} When $j$ becomes
larger, the loss is smaller. For the spin enhancement ($\Delta j=-1$),
we can develop a similar theoretical basis. So far, we have only
considered $\Delta j=\pm 1$.  $\Delta j=0$ only occurs for the
transition between one $\psi^a$ and one $\psi^b$ state, and leads to
the absolute spin reversal.  There are two channels, $\psi_a
\rightarrow \psi_b$ and $\psi_b \rightarrow \psi_a$.  Transitioning
from $\psi_a$ to $\psi_b$ switches spin from up to down, while
transitioning from $\psi_b$ to $\psi_a$ switches spin from down to
up. This simple picture nicely explains why LP creates multiple
domains with mixed spin up and spin down.\cite{stanciu} {
  The final outcome, whether it is demagnetization, magnetization or
  spin reversal, critically depends which transitions dominate. }

For left (LC) and right (RC) circularly polarized light, the situation
is more complicated since now $m_j$ plays a role. This is reflected in
Table \ref{table1}. However, we find that there is a simple and
similar dependence of $\eta$ on $j$ as LP. If we ignore those $m_j$
terms in $\eta$, each of the resultant $j$ terms under LC and RC is
exactly the same as LP, comparing column 2 with 3 and column 2 with
4. The magnetic quantum number $m_j$ opens new channels to manipulate
spin. For instance, in the second column where the transition is
between $\psi^a$ states and $\Delta j=+1$, if $m_j=-1/2$, for LC,
$\eta=-j/(j+1)<0$, corresponding to spin reversal. The same is true
for $\Delta j=-1$. In fact, all the six channels are open for spin
reversal, in comparison to two channels in LP; the same can be said
for magnetization and demagnetization.  This explains why LC and RC
appear more powerful to switch spins than LP.\cite{sabine2012} For RC,
the situation is similar if we have $m_j=1/2$.  Our results question
again whether it is appropriate to label the helicity-independent
switching as thermal switching, since the entire process is still
optical.\cite{jpcm13}

In the following, we will construct a model to demonstrate that the
insight gained from these spin-orbit coupled states is useful and
appears in our calculation below.

\subsection{Spin reversal theory: cw limit}

In contrast to the title of this subsection, our original idea was to
derive an analytic expression for traditional magneto-optics under
cw excitation from a simple model suggested by Bigot\footnote{Private
  conversation}  to
us in 1999. Bigot wondered whether it is possible to develop a simple
model to compare the theory with the experimental finding. Note,
however, that Pershan \ete\cite{pershan} and others\cite{bene} already
used the simple oscillator model to compute the magneto-optical
response. Common to these theories is that the magnetic field is used,
since the classical theory traditionally does not have spin in it. A
nice feature of such an approach is that an analytic solution for
the diagonal and off-diagonal susceptibilities is possible.

However, we have no prior bias toward the magnetic field, since our prior
theoretical investigations\cite{prl00,jap08,np09} often do not have a
magnetic field. So, we simply replaced the magnetic field by the
spin-orbit coupling,\cite{jpcm13,mingsu10,jpcm11,jpcm14,jpcm15,epl15}
\be H=\frac{{\bf p}^2}{2m}+\frac{1}{2}m\Omega^2 {\bf r}^2 +\lambda
    {\bf L}\cdot {\bf S} -e {\bf E}(t) \cdot {\bf r} \label{ham1}. \ee
    Here, the first term is the kinetic energy operator of the
    electron; the second term is the harmonic potential energy
    operator with system frequency $\Omega$; $\lambda$ is the
    spin-orbit coupling in units of eV/$\hbar^2$; $ {\bf L}$ and $
    {\bf S} $ are the orbital and spin angular momenta in units of
    $\hbar$, respectively, and {\bf p} and {\bf r} are the momentum
    and position operators of the electron, respectively. Note that
    ${\bf L}$ is computed from ${\bf L}={\bf r}\times {\bf p}$, and
    there is no need to set up a different equation for it.

The entire equation of motion can be written down as\cite{epl15} \ba \frac{d\bf r}{dt}&=&\frac{\bf p}{m}-\lambda
({\bf r}\times {\bf S})\label{position},\\ \frac{d \bf
  p}{dt}&=&-m\Omega^2 {\bf r}+e{\bf E}(t) -\lambda {\bf p}\times {\bf
  S} \label{momentum},\\ \frac{d\bf S}{dt}&=&\lambda ({\bf L} \times
{\bf S})\label{spin},\\ \frac{d\bf L}{dt}&=&-e{\bf E}(t) \times {\bf
  r} - \lambda ({\bf L} \times {\bf S}), \label{orbital}\ea where the
last equation does not enter the calculation and is left here for
later usage. The total angular momentum ${\bf J}$ is determined by
the laser field and position vector.  If we assume that the spin is
constant, then three coupled equations (\ref{position})-(\ref{spin})
can be simplified as\cite{epl15} \be \ddot{\bf r}+2\lambda \dot{\bf
  r}\times {\bf S}+ (\Omega^2-\lambda^2S^2){\bf r}-\lambda^2({\bf
  r}\cdot{\bf S}){\bf S}=\frac{e{\bf E}(t)}{m}.\label{new} \ee If the
external field is cw, we can derive the susceptibilities
analytically as \ba \chi_{xx}^{(1)}(\omega)&=&-\frac{Ne^2}{\epsilon_0
  m}\frac{\Omega^2-\omega^2-\lambda^2S_z^2 }{
  (\Omega^2-\omega^2-\lambda^2S_z^2)^2-(2\lambda
  S_z\omega)^2}\label{chi1}
\\ \chi_{xy}^{(1)}(\omega)&=&-i\frac{Ne^2}{\epsilon_0 m}\frac{2\lambda
  S_z\omega}{(\Omega^2-\omega^2-\lambda^2S_z^2)^2-(2\lambda
  S_z\omega)^2}, \label{xy}\ea where $N$ is the number density and
$\epsilon_0$ is the permittivity in vacuum. This result convinces
us that the model can describe the basic features of the
magneto-optics.

\subsection{Spin reversal theory: pulsed laser}

With the success of our model to describe the basic magneto-optics, we
wonder whether such an equation allows us to describe spin switching.  It
is easy to show from Eq. (\ref{spin}) that the spin is conserved, and
its module does not change with time. This indicates that spin
switching may be possible.

The key step has been outlined in our study.\cite{epl15} Here, we give a
summarized account.  To start with, we solve the equations of motion
numerically.  We choose laser pulses of two different kinds.  For a
linearly polarized ($\pi$) pulse, the electric field is $ {\bf
  E}(t)=A_0 {\rm e}^{-t^2/\tau^2}\cos(\omega t) \hat{x},$ where
$\omega$ is the laser carrier frequency, $\tau$ is the laser pulse
duration, $A_0$ is the laser field amplitude, $t$ is time, and
$\hat{x}$ is the unit vector along the $x$ axis. Note that the results
are the same if the field is along the $y$ axis. The electric field
for the right- and left-circularly polarized pulses ($\sigma^{+}$ and
$\sigma^{-}$) is $ {\bf E}(t)= A_0 {\rm e}^{-t^2/\tau^2} (\pm
\sin(\omega t) \hat{x}+\cos(\omega t) \hat{y})$, where $+(-)$ refers
to $\sigma^{+}$($\sigma^{-}$).  We then compute the spin evolution by
numerically solving the three coupled equations
(\ref{position}-\ref{spin}). Since we did not know whether the spin
momentum has any major effect on the spin reversal, we choose the
initial spin momentum $S_z(0)=2.2\hbar$. It turns out that this is a
crucial step. Had we chosen a smaller value, we could miss the spin
reversal entirely since too small a spin could not reverse the spin,
regardless of the laser field amplitude. We choose the spin-orbit coupling
$\lambda=0.06{\rm eV}/\hbar^2$, and $\hbar\omega=\hbar\Omega=1.6$
eV. This is a resonant excitation. If we have an off-resonant
excitation, the spin can not be switched over effectively. For this
reason, our theory is based on photon absorption, or
photomagnetism. Our pulse duration is $\tau=60$ fs and amplitude is
0.035 $\rm V/\AA$, which is already optimized. We also assume that the
electron moves along the $z$ axis with velocity 1 $\rm \AA$/fs or
10$^5$ m/s, which is slightly lower than the Fermi velocity. We find
that this initial velocity is necessary due to the uncertainty
principle; otherwise, since the laser field is only in the $xy$ plane,
the electron only moves in the $xy$ plane, and there is no way to have
a nonzero orbital angular momentum along the $x$ or $y$ axis, so the
term on the left-hand side of Eq. (\ref{spin}) is zero. A larger value
of $v_z$ leads to a larger effect in the spin reversal. Here, we want
to be conservative, so we choose a smaller value. The initial values
of $v_x$ and $v_y$ do not matter too much since the laser is in the
$xy$ plane anyway.

\begin{figure}[th]
\centerline{\psfig{file=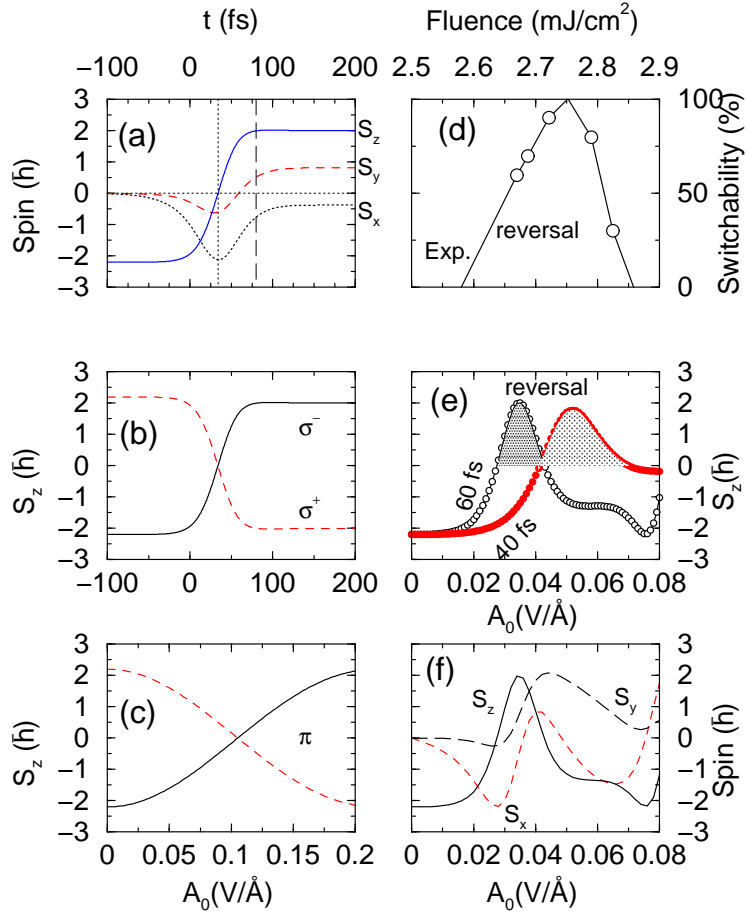,angle=270,width=10cm}}
\vspace*{8pt}
\caption{ (a) All-optical spin reversal for $S_x$, $S_y$ and $S_z$ as
  a function of time $t$.  The vertical dotted line denotes the time
  when $S_z$ passes through zero, and the vertical dashed line denotes
  the time when the spin reversal starts. Here $\sigma^{-}$ is used,
  the laser pulse duration is $\tau=60$ fs, and the field amplitude is
  0.035 $\rm V/\AA$.  (b) The $\sigma^{-}$ pulse (solid line) only
  switches spin from down to up, while the $\sigma^{+}$ pulse (dashed
  line) only switches spin from up to down. (c) The $\pi$ pulse can
  switch spin from up to down or from down to up, but at a much higher
  field amplitude.  (d) Experimental spin reversal window from
  Ref. 28.  (e) Final spin angular momentum $S_z$ as a function of the
  laser field amplitude for laser duration $\tau=60$ fs (empty
  circles) and 40 fs (filled red circles). The shaded regions are the
  spin reversal window.  (f) As the field amplitude increases, the
  spin angular momentum changes from non-switching, canting along the
  $-x$ axis, switching, and canting along the $+y$ axis.  The figure
  is from Ref. 112 used with permission from EPL.}
\label{fig1}
\end{figure}

 Figure \ref{fig1}(a) shows the spin reversal on this single site. We
 initialize the spin along the $-z$ axis.  Upon the laser excitation,
 the spin first precesses strongly without any oscillation toward the
 $xy$ plane, and the exact precession of $S_x$ and $S_y$ depends on the
 initial phase of the laser pulse,\footnote{Our current integration
   starts at $-1000$ fs, where the amplitude of the laser field is
   extremely small.  If we change it to a different time, then $S_x$
   and $S_y$ may behave differently, depending on the initial phase
   that the laser acquires.  However, $S_z$ remains the same.}
 but the precession of $S_z$ is always the same.  The spin reaches the
 negative maximum at 34 fs, exactly when $S_z$ passes through
 zero. $S_z$ successfully switches to $2\hbar$ at 80 fs, where the
 spin rotates 155.9$^\circ$.  Figure \ref{fig1}(b) shows that both
 $\sigma^{+}$ and $\sigma^{-}$ can switch spin within a few hundred
 femtoseconds.  Our results reveal a stringent symmetry constraint on
 the spin switching: the $\sigma^{-}$ light only switches the spin
 from down to up, while the $\sigma^{+}$ light switches the spin from
 up to down, not the other way around. We have also tested the linearly
 polarized light. We find that depending on the initial spin
 orientation, LP can switch from down to up and from up to down (see
 Fig. \ref{fig1}(c)).  More importantly, the needed laser amplitude is
 much larger.  To induce spin switching, we need to increase
 $A_0$ above 0.2$\rm V/\AA$, or 5.7 times higher than that used for either
 $\sigma^{+}$ or $\sigma^{-}$. This result is consistent with the
 finding by Alebrand \ete\cite{sabine2012}

 Vahaplar \ete\cite{vahaplarprl} found that AOS does not occur with any
 laser field fluence. They discovered that the spin reversal window is
 very narrow and asymmetric (see Fig. \ref{fig1}(d)). Our theoretical
 results are shown in Fig.  \ref{fig1}(e). We see that as $A_0$
 increases, the final spin $S_z$ first increases sharply (see the
 empty circles) and then reaches its maximum of $2\hbar$ at
 $A_0=0.035$$\rm V/\AA$, where the spin is reversed.  If we increase
 the field amplitude further, $S_z$ decreases, and eventually the spin
 switching disappears.  The reversal window is indeed very narrow and
 asymmetric (see the shaded region in Fig. \ref{fig1}(b)), only from
 0.026 to 0.042 $\rm V/\AA$.

It is interesting to investigate how the spin changes for those
unoptimized amplitudes. For this, we plot all three components of
the spin as a function of the laser field amplitude. It is clear that
when the amplitude is small, the spin change is small. But as we
gradually increase it to about 0.026 $\rm V/\AA$, the final spin
simply cants along the $-x$ axis. On the other hand, if we have too
big an amplitude above 0.035 $\rm V/\AA$ but below 0.06 $\rm V/\AA$,
the spin cants to the $+y$ axis. Therefore, a competition between spin
canting and spin reversal leads to the asymmetric dependence of the
switchability on the laser amplitude. We find a better agreement with
the experimental one\cite{vahaplarprl} than their own theoretical
results. This is the first indication that our theory may catch
something important here.

\subsection{Exchange interaction and Rise of inverse Faraday effect}

However, Vahaplar \ete\cite{vahaplarprl} included the exchange
interaction between different spins and their system is much larger
than ours. In our case, we basically have a single site.  To properly
include the exchange coupling, we construct the following Hamiltonian
\cite{jpcm11,jpcm13} \be H=\sum_i \left [\frac{{\bf p}_i^2}{2m}+V({\bf
    r}_i) +\lambda {\bf L}_i\cdot {\bf S}_i -e {\bf E}({\bf r}, t)
  \cdot {\bf r}_i\right ]-\sum_{ij}J_{ex}{\bf S}_i\cdot {\bf
  S}_{j}. \label{ham} \ee The summation is over all the lattice sites.
Here, the first four terms are the same as our Hamiltonian
(\ref{ham1}).  The last term is the exchange interaction, and $J_{ex}$
is the exchange integral in units of eV/$\hbar^2$.  Such a Hamiltonian
contains the necessary ingredients for AOS; a similar form is often
used for magnetic multilayers.\cite{hs,stiles2}

We consider a slab of $101\times 101 \times 4$ lattice sites, where we
can shed some new light on the inverse Faraday effect.  We take
$S(0)_z=1.2\hbar$ as an example. The field amplitude is at its optimal
value of $A_0=0.018~\rm V/\AA$.  As stated above, the orbital angular
momentum evolves with time according to \be \frac{{d\bf L}_i} {dt}=
-e{\bf E}(t) \times {\bf r}_i - \lambda ({\bf L}_i \times {\bf
  S}_i). \label{oo} \ee The two driving terms on the right-hand side
represent two torques.  The first term is the torque due to the laser
field, $\tau_{laser}=-e{\bf E}(t) \times {\bf r}_i$. { Note that here
$\tau$ is the torque, not to be confused with the duration above.} This is a
cross-product of the laser field and electron position, not that of
the field and itself, which is in sharp contrast to Eq. (\ref{hom})
under cw excitation. Our finding is also consistent with the study by
Popova {\it et al.},\cite{pop} where the effective field is not a
simple product of two fields. Different from Popova \ete, our
effective field acts upon the orbital angular momentum, not spin. If
we directly compute the spin momentum change, the net change in the
module is small, but the spin precesses very strongly.

\begin{figure}
\centerline{\psfig{file=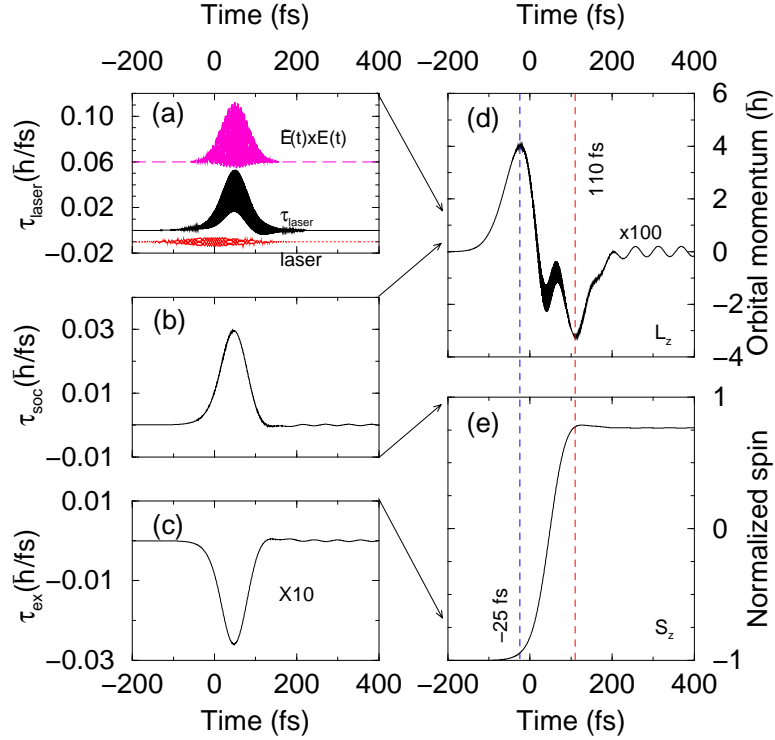,width=10cm,angle=270}}
\vspace*{8pt}
\caption{ (a) Laser-induced torque $\tau_{laser}$ as a function of
  time. The peak is at 50 fs, different from the laser field (see the
  bottom curve) peak at 0 fs. The long-dashed line refers to the
  effective field computed from ${\bf \tilde{E}}(t) \times {\bf
    E}(t)$.  Here $S_z(0)=1.2\hbar$.  (b) Spin-orbit torque change
  $\tau_{soc}$ with time. This is the main torque to reverse the
  spin. (c) Exchange-interaction torque change $\tau_{ex}$ with
  time. It is at least 10 times smaller than the spin-orbit torque.
  (d) Orbital angular momentum $L_z$ as a function of time. It peaks
  at -25 fs, which is ahead of spin reversal by 135 fs.  The first
  vertical dashed line refers to the peak time.  (e) Spin reversal.
  The spin is initialized along the $-z$ direction, and upon laser
  excitation, the spin is reversed to the $+z$ direction around 110
  fs.  The spin is normalized with respect to the initial spin
  $S_z(0)=1.2\hbar$.  }
\label{supfig1}
\end{figure}

The solid curve in Fig. \ref{supfig1}(a) shows
the $z$ component of $\tau_{laser}$, while the dotted curve is our
laser E-field. It is clear that $\tau_{laser}$ is quite
different from the laser pulse itself: (i) it peaks at 50 fs, not 0
fs; (ii) it is asymmetric; and (iii) its values are mostly positive.
However, if we assume that the position behaves like $|{\bf
  \tilde{E}}(t)| \propto \cos(\omega t +\pi/5)
\exp(-(t-54)^2/64.7^2)$, $\tau_{laser}$ can be reproduced to some
extent by ${\bf \tilde{E}}(t) \times {\bf E}(t)$ (see the long dashed
line). This reminds us of the inverse Faraday
effect,\cite{rasingreview} but this cross product provides a source
term for the electron orbital angular momentum, not for the
spin.\cite{stanciu,vahaplarprb,ostler} From Eq. (\ref{oo}), we see
that $\tau_{laser}$ is not the only torque that affects the
orbital. Once the orbital angular momentum differs from zero,
$\tau_{soc}=-\lambda {\bf L}_i\times {\bf S}_i$ builds up and starts
to contribute a negative torque to the orbital angular momentum (see
Fig. \ref{supfig1}(b)).  Figure \ref{supfig1}(c) shows that the
exchange torque $\tau_{ex}$ is at least one order of magnitude weaker
than $\tau_{soc}$, so $\tau_{soc}$ dominates over $\tau_{ex}$.

Figure \ref{supfig1}(d) shows that upon laser excitation, the orbital
angular momentum $L_z$ in the first layer of atomic sites increases
sharply from 0.0$\hbar$ to 0.04$\hbar$ at -25 fs, after which it
swings to the negative maximum of -0.03$\hbar$ before it returns to
zero and oscillates around it. Other layers have a similar dependence
(not shown).  A competition between $\tau_{laser}$ and $\tau_{soc}$
leads to a sudden reduction of $L_z$ around 18 fs and subsequent
reversal at 110 fs, which explains the violent oscillation in $L_z$
(see Fig. \ref{supfig1}(d)).  The change in spin is quite
different. Figure \ref{supfig1}(e) shows that the normalized spin starts
from -1 and flips over to 0.75 at 110 fs. The spin dynamics delays
with respect to the orbital dynamics by 135 fs.  The peak of
$\tau_{soc}$ corresponds to the maximum change in $S_z$, while the
small value of $\tau_{soc}$ after the peak ensures that the spin can
not be switched back.  $\tau_{soc}$ provides a necessary positive
torque that finally flips the spin from the $-z$ to $+z$ direction.

\subsection{Effect of the laser field amplitude on spin reversal}

\begin{figure}
\centerline{\psfig{file=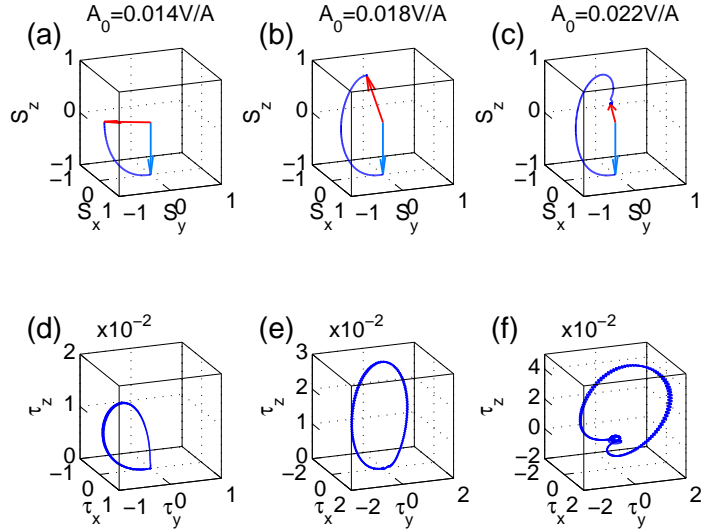,width=11cm,angle=0}}
\vspace*{8pt}
\caption{ Effect of the laser field amplitude on spin reversal.
  (a) Three-dimensional spin precession for the field amplitude
  $A_0=\rm 0.014~V/\AA$.  Here $S_z(0)=1.2\hbar$.  The spin is initiated
  along the $-z$ axis (see the blue arrow). The red arrow denotes the
  final spin. All the spins are normalized.  (b) Same as (a) but with
  field amplitude $A_0=\rm 0.018~V/\AA$. Here the spin does reverse
  successfully.  (c) Same as (a) but with field amplitude
  $A_0=\rm 0.022~V/\AA$. Here the spin passes the optimal location.  (d)
  Three-dimensional plot of the spin-orbit torque $\tau_{soc}$ for
  $A_0=0.014~\rm V/\AA$. It is mostly positive but small. All the torques
  are multiplied by 100 for easy viewing and are in the units of
  $\hbar$/fs.  (e) Same as (d), but with $A_0=0.018~\rm V/\AA$. This is the
  case that the torque is big and positive. (f) Same as (d), but with
  $A_0=0.022~\rm V/\AA$. In the beginning of the laser excitation, this
  torque is big and positive, but at the end, becomes negative. This
  forces the spin  away from the $+z$ axis.  }
\label{supfig2}
\end{figure}

We have already seen how the laser field amplitude affects spin
reversal at a single site. Here we have a much larger system.  To
reveal further insight into the effect of the laser field amplitude on
spin reversal, we choose three amplitudes, $A_0=0.014$, $0.018$ and
$0.022~\rm V/\AA$, and $S_z(0)=1.2\hbar$.  We take the first layer as
an example, since the rest of them behave similarly except for a
slight time delay. The spin configuration is initialized along the
$-z$ axis (see the light blue arrows in Figs. \ref{supfig2}(a-c)). We
start with $A_0=0.014~\rm V/\AA$.  We find that as time evolves, the
spin does try to flip, but the final spin only tilts toward the $yz$
plane, and it does not have enough power to reach the full reversed
spin configuration.  If we increase the amplitude to $A_0=0.018~\rm
V/\AA$, we see from Fig. \ref{supfig2}(b) that the spin is capable of
flipping into the opposite direction. Now if we increase the laser
amplitude further to $A_0=0.022~\rm V/\AA$, the situation changes.
Figure \ref{supfig2}(c) shows that the spin rotates too much; and once
the field is off, it cants toward the $xy$ plane.

The above field dependence can be understood from the spin-orbit
torque. Figures \ref{supfig2}(d-f) show the respective three-dimensional
torques for each amplitude. Note that all the torques in the figure
are in the units of $\hbar$/fs. Since our initial spin is in the $-z$
direction and our desired final spin is in the $+z$ direction, the
torque has to be positive and must be dominated by the $z$ component;
otherwise, we only see the spin canting, not switching.  Figure
\ref{supfig2}(d) shows that the torque starts from 0 and gradually
grows. It is positive, but the amplitude is small only around 0.01
$\hbar$/fs, which explains why the spin can not be switched over. At
$A_0=0.018~\rm V/\AA$, the torque is positive and larger than that at
$A_0=0.014~\rm V/\AA$. This is the origin of the spin reversal. If the
field amplitude is increased to $A_0=0.022~\rm V/\AA$, the situation
in the earlier stage is similar to that at $A_0=0.018~\rm V/\AA$, but
since the amplitude is larger, the torque is larger. As time evolves, it precedes to a negative value of -0.02 $\hbar$/fs. This proves
to be detrimental to the spin reversal, which explains why the final
spin passes the optimal configuration. To summarize, to reverse spins,
the laser field amplitude has to be in the narrow region, where the
spin-orbit torque is not too small, and not too big to turn
negative. This competition leads to a narrow region, as seen in
experiments.\cite{vahaplarprb}

\section{New techniques and new directions}

He \ete\cite{he} demonstrated AOS using a telecom-band femtosecond
fiber laser, and then they fabricated Hall cross devices that can read
out the AOS signal by measuring the anomalous Hall voltage
changes. Such an incredible technique becomes very useful to
characterize AOS switching.\cite{hadri,elhadri} The amorphous TbFeCo
thin films also show bistable magneto-resistance
states.\cite{li} Such an exchange bias device can perform even at room
temperature.

El Hadri \ete\cite{elhadri} directly applied the ferromagnetic Hall
crosses\cite{he} to characterize the AOS in ferromagnetic Pt/Co/Pt
heterostructures. AOS is measured through the anomalous Hall
effect. Such a technique provides an alternative to the magnetic
imaging technique. They found that the AOS in Pt/Co/Pt is a cumulative
process, where a certain number of optical pulses is necessary to
obtain full and reproducible switching.  This finding was also found
in granular FePt by Takahashi \ete\cite{takahashi} This is quite
different from AOS in GdFeCo.\cite{hadri}  They suggested this could
become a new type of opto-spintronic device. Such an idea could be
incorporated with electric switching, such as the recently
demonstrated switching in CuMnAs.\cite{wadley}

In 2015, Satoh \ete\cite{satoh} demonstrated that an arbitrary optical
polarization write/read is possible in antiferromagnet YMnO$_3$. This
idea is consistent with the theoretical prediction by Gomez-Abal
\ete\cite{go} However, the entire process is a rapid beating and does
not settle down to a definitive state. This potentially will limit its
application, if used as a switching device.

Ogawa \ete\cite{ogawa} studied the local excitation of the spin in
magnetic skyrmions Cu$_2$OSeO$_3$. They showed that the spins in the
conical and skyrmions phases can be excited by the effective impulsive
magnetic field from the inverse-Faraday effect.  It would be
interesting if the spin could be permanently switched.

Very recently, Goncalves \ete \cite{gon} demonstrated that a sub-10-fs
pulse is able to efficiently excite a magnetic system such as
GdFeCo. This revealed unprecedented details on the electron-electron
interaction time scale. Note that all the previous investigations were
carried out over tens of femtoseconds.\cite{stanciu} It is likely that
the observed picosecond spin reversal in fact starts earlier. It is
the  current magnetic image techniques that limit our view on the
shortest possible time scale.

\section{Conclusions}

We have presented a short review on all-optical spin switching and a
simple model that works quite well to explain all-optical
helicity-dependent spin switching (AOS). To give the reader a complete
picture of AOS, we have discussed the initial phase of femtomagnetism
with emphasis on ultrafast demagnetization. Then we focus on the
experimental discovery of AOS as a major branch of femtomagnetism. AOS
is an extremely active research area at this time. The new theoretical
and experimental findings are reported frequently. It is impossible to
include all aspects of the current research, so we limit our focus to
four major directions of the experimental discoveries: (1) temperature
effect; (2) laser parameter effect; (3) effect of the sample
composition; and (4) AOS samples beyond GdFeCo.

Before we turn to our latest research on AOS at a single spin site, we
show that a simple all-optical spin switching rule exists, and it
provides essential guidance as to how microscopically the light reverses
the spin and how the light helicity affects the AOS. We find that
although our spin-orbit coupled harmonic model is very simple, it
contains some important elements for magneto-optics and all-optical
spin switching. Even in the static limit, we show that the simple
model reproduces the well-known dependence of the diagonal and
off-diagonal susceptibility on the spin angular momentum and
spin-orbit coupling. We show numerically that spin reversal is
possible within such a model. The results are very good. For instance,
they match the experimental laser fluence dependence, even better than
other more complicated models. Even more interesting is that our model
reveals that the spin-orbit torque plays the role as an effective
magnetic field, an equivalent field to the inverse Faraday effect,
which has long been sought after.

In summary, additional experimental and theoretical work is necessary
to iron out the details of the complicated underlying mechanism of the
all-optical switching. On the theoretical side, at present, the
theoretical efforts largely follow the experimental development. There
have been very few predictions, partly because the majority of the
theories are heavily phenomenological and empirical. There is a need
to include true laser pulses, not heat pulse or effective magnetic
field. Much fewer studies are in the structural study. This is a new
direction that is going to be important in the future.
Experimentally, pursuing a simpler system, as a way to understanding
the complex rare-earth compounds, is very helpful. This should be
coupled with the element-specific technique, so one can be sure which
spin and which element is switched. A systematic investigation of the
composition is needed.  We think that this is the single most
important addition to AOS. A joint effort between theory and
experiment is expected to yield new and more exciting results, and
possibly open new applications in femtomagnetism.

\section*{Acknowledgments}

 This work was solely supported by the U.S. Department
of Energy under Contract No. DE-FG02-06ER46304. Part of the work was
done on Indiana State University's quantum cluster and
high-performance computers.  The research used resources of the
National Energy Research Scientific Computing Center, which is
supported by the Office of Science of the U.S. Department of Energy
under Contract No. DE-AC02-05CH11231. This work was performed, in
part, at the Center for Integrated Nanotechnologies, an Office of
Science User Facility operated for the U.S. Department of Energy (DOE)
Office of Science by Los Alamos National Laboratory (Contract
DE-AC52-06NA25396) and Sandia National Laboratories (Contract
DE-AC04-94AL85000). We thank Dr. A. Kimel for sending us some useful
remarks after this paper was published.

\end{document}